\documentclass[conference, a4paper]{llncs}
\usepackage{comment}
\usepackage{cite}
\usepackage{amsmath,amssymb,amsfonts}
\usepackage{graphicx}
\usepackage{textcomp}
\usepackage{xcolor}
\usepackage{pgfplots}
\pgfplotsset{compat=1.17}
\usepackage{tikz}
\usetikzlibrary{patterns}
\usepackage{algorithm2e}
\usepackage{caption}
\usetikzlibrary{patterns}
\usepackage{subcaption}

\def\BibTeX{{\rm B\kern-.05em{\sc i\kern-.025em b}\kern-.08em
    T\kern-.1667em\lower.7ex\hbox{E}\kern-.125emX}}

\begin{document}

\title{Privacy-Preserving Credit Card Approval Using Homomorphic SVM: Toward Secure Inference in FinTech Applications}

\author{Faneela\inst{1} \and Baraq Ghaleb\inst{1} \and Jawad Ahmad\inst{2} \and William J. Buchanan\inst{1} \and Sana Ullah Jan\inst{1}
} 

\author{
Faneela\inst{1}, Baraq Ghaleb\inst{1}, Jawad Ahmad\inst{2}, William J. Buchanan\inst{1}, Sana Ullah Jan\inst{1}
}

\institute{
Blockpass ID Lab, Edinburgh Napier University, Edinburgh.
\and
College of Business Administration, Prince Mohammad Bin Fahd University, Al Khobar, Saudi Arabia.
}

\maketitle
\begin{abstract}
The growing use of machine learning in cloud environments raises critical concerns about data security and privacy, especially in finance. Fully Homomorphic Encryption (FHE) offers a solution by enabling computations on encrypted data, but its high computational cost limits practicality. In this paper, we propose PP-FinTech, a privacy-preserving scheme for financial applications that employs a CKKS-based encrypted soft-margin SVM, enhanced with a hybrid kernel for modeling non-linear patterns and an adaptive thresholding mechanism for robust encrypted classification. Experiments on the Credit Card Approval dataset demonstrate comparable performance to the plaintext models, highlighting PP-FinTech’s ability to balance privacy, and efficiency in secure financial ML systems.
\end{abstract}


\section{Introduction}
The rise of cloud computing has accelerated machine learning adoption in sensitive sectors like finance and healthcare. While offering scalability and efficiency, outsourcing ML tasks to the cloud raises critical concerns about data privacy and result integrity ~\cite{sahu2024homomorphic}. Indeed, numerous ML-based financial systems have emerged, with Support Vector Machines (SVMs) gaining popularity due to their effectiveness in classification tasks~\cite{jana2024mathematical}, but their reliance on external computation risks exposing sensitive data and receiving unverified outputs. To address these challenges, we propose PP-FinTech, a privacy-preserving SVM scheme leveraging the CKKS fully homomorphic encryption scheme, which supports real-valued encrypted computation ~\cite{lee2023configurable}. Our approach addresses both data confidentiality and computational correctness, two key challenges in secure financial ML systems~\cite{marcolla2022survey}. The key contributions of this paper are:

\begin{itemize}
    \item A CKKS-encrypted SVM with a hybrid Polynomial–RBF kernel for modeling non-linear data.
    \item An adaptive thresholding method to improve encrypted classification robustness under noise.
    \item An optimized inference pipeline using SIMD (Single Instruction, Multiple Data) techniques, allowing multiple encrypted samples to be processed in parallel and reducing per-sample latency.
    \item A careful selection of encryption parameters to stay within the noise budget for reliable decryption.
    \item Experimental evaluation on the Credit Card Approval dataset showing promising performance while ensuring strong privacy guarantees.
\end{itemize}

\section{Preliminaries}
\subsection{Support Vector Machine (SVM)}

Support Vector Machine (SVM) is a supervised learning algorithm widely used for classification tasks, particularly effective for linearly separable problems. Its primary objective is to identify the optimal hyperplane that maximizes the margin between different classes in the training dataset. For cases involving non-linear data distributions, SVM leverages kernel functions—such as polynomial and Gaussian (RBF) kernels—to implicitly map the input data into a higher-dimensional feature space, where linear separation becomes feasible~\cite{wu2021power}. The classification decision function of an SVM can be expressed as:
\[
\hat{y} = \text{sign}\left(\sum_{i \in S} \alpha_i y_i K(x_i, t) + b\right)
\]

Here, $S$ denotes the set of indices corresponding to the support vectors, while $x_i$ and $y_i$ represent the support vectors and their associated labels, determined during the training phase. The coefficient $\alpha_i$ is the Lagrange multiplier for each support vector, $t$ is the input test sample, and $b$ is the bias term. The function $K(x_i, t)$ is the kernel function that measures the similarity between the training sample $x_i$ and the test sample $t$. The output $\hat{y}$ represents the predicted class label. 

\subsection{CKKS Scheme}

The CKKS scheme \cite{cheon2017homomorphic} is a leveled fully homomorphic encryption framework based on the Ring Learning With Errors (RLWE) problem, designed to support approximate arithmetic on encrypted real-valued data~\cite{kim2022approximate}. It operates with key parameters such as the ring dimension \( n \), multiplicative depth \( L \), modulus chain \( q_L \geq \ldots \geq q_1 \), and scaling factor \( \Delta \), which together control the precision and noise growth during encrypted computations. CKKS also supports SIMD encoding, allowing multiple values to be packed into a single ciphertext. This property enables efficient batch processing during inference, which we leverage in our encrypted SVM pipeline. Further details about parameter settings and noise budget control are described in the secure inference section.

In our implementation, we employed the OpenFHE library to perform all homomorphic operations using the CKKS scheme. OpenFHE is an open-source fully homomorphic encryption library that provides implementations of state-of-the-art FHE schemes ~\cite{OpenFHE}.

\section{Related Work}
Encrypted inference using SVMs has been the subject of increasing attention in privacy-preserving ML. For instance,  Park et al.\cite{park2022privacy} proposed an SVM framework built on Homomorphic Encryption (HE) that incorporates fairness constraints during training. Deng et al.\cite{deng2022privacy} introduced a soft-margin SVM model tailored for secure disease diagnosis. Other studies, including~\cite{xie2021achieving} and~\cite{hu2023achieving}, have developed privacy-preserving diagnosis systems and verifiable SVM classification protocols across multiple clouds using HE. While these approaches demonstrate secure inference capabilities, they either operate under partially encrypted settings or lack support for practical functionalities such as batch processing, noise tolerance, and encrypted post-processing
Beyond SVMs, HE has also been employed in other machine learning models, such as logistic regression~\cite{naresh2025exploring} and deep neural networks~\cite{naresh2024ppdnn, sirisha2023privacy}, particularly in healthcare and credit scoring contexts. However, these works primarily focus on linear or deep learning-based models, without exploring encrypted non-linear classification using SVMs.
In contrast to prior work, our approach introduces a fully encrypted inference pipeline using a hybrid-kernel SVM implemented under the CKKS scheme. We incorporate adaptive thresholding to mitigate the effects of encryption-induced noise and leverage SIMD batching to enhance runtime performance. This work bridges a key gap between cryptographic privacy guarantees and practical, real-time encrypted inference, particularly in financial application domains.

\section{Proposed Scheme}
We propose PP-FinTech, a privacy-preserving SVM classification framework based on the CKKS FHE scheme. The approach consists of two phases: plaintext model training and encrypted inference. During training, an SVM is trained on an 80:20 split of the Credit Card Approval dataset, producing support vectors, dual coefficients, and bias. These parameters are encrypted using CKKS to enable secure inference over encrypted input data. Predictions remain encrypted during computation and are decrypted only by the user, ensuring complete data privacy. The proposed model is detailed in the following subsections:

\subsection{Model Training and Feature Selection }

In the training phase, a standard SVM model is trained on plaintext data using the Credit Card Approval dataset, which is which underwent multiple preprocessing and feature selection stages.  To ensure the robustness and effectiveness of the proposed SVM model, we conducted a thorough preprocessing pipeline, encompassing normalization and feature selection techniques tailored for financial datasets. The dataset employed for training and evaluation is the Credit Card Approval dataset, obtained from the UCI ML Repository. This dataset comprises 690 instances and 15 attributes, a mix of both numerical and categorical features such as age, income, employment status, and credit history indicators. Due to its real-world origin and widespread use in benchmarking financial decision systems, it serves as an ideal candidate for evaluating the performance of privacy-preserving classification models.
Before training the model, all numerical attributes are normalized using the \texttt{StandardScaler} method, a widely accepted technique for standardizing features by removing the mean and scaling to unit variance. This step is essential for mitigating issues arising from feature dominance and ensuring numerical stability, particularly when using distance-based classifiers such as SVMs ~\cite{razavi2023machine}. The transformation is mathematically defined as:
\[
X' = \frac{X - \mu}{\sigma}
\]

where \( X \) denotes the original feature value, \( \mu \) is the mean, and \( \sigma \) is the standard deviation. By centering and scaling the input features, we ensure that each contributes equally to the model’s decision boundary, preventing any bias caused by differing value ranges.

Following normalization, a filter-based feature selection approach is employed to reduce the dimensionality of the dataset and improve computational efficiency during encrypted inference. Filter methods assess the relevance of each feature independently based on statistical measures such as correlation with the target class. This process aids in eliminating redundant or irrelevant features, which can otherwise introduce noise and degrade model performance especially in HE settings, where computational overhead is sensitive to input size. The selection process identified a subset of features that retained strong predictive power while significantly reducing the dimensionality of the input space. 

\subsubsection{Hybrid Kernel Approach}

To improve classification accuracy and generalization, we adopt a hybrid kernel combining Polynomial and Radial Basis Function (RBF) kernels. The polynomial kernel captures high-order feature interactions, while the RBF kernel models local, non-linear patterns by projecting inputs into an infinite-dimensional space, allowing the model to form complex and flexible decision boundaries even with limited training data ~\cite{pande2023comparative}. This combination leverages the strengths of both kernels, enabling the model to handle diverse data structures without substantial computational overhead. The hybrid kernel function is mathematically defined as:
\[
K(X', SV_j) = \lambda_1 K_{p}(X', SV_j) + \lambda_2 K_{R}(X', SV_j)
\]
where \( \lambda_1 \) and \( \lambda_2 \) are weight factors controlling the contribution of each kernel. In our implementation, the weight parameters were tuned using a grid search approach on a validation set. The best empirical results were obtained with \( \lambda_1 \) = 0.7 and \( \lambda_2 \) = 0.3. This configuration achieved an optimal balance between classification accuracy, model interpretability, and generalization performance. Additionally, we adopt a soft-margin SVM, which introduces slack variables to balance margin maximization with allowable misclassifications. This flexibility is essential for noisy or overlapping classes, as often seen in real-world financial data where perfect separability is rarely possible.
\vspace*{5mm}

\subsubsection{Conservative Scaling for Error Recovery}
To ensure stable training, we apply a conservative scaling strategy~\cite{krishnamurthy2022efficient} that adjusts model updates based on gradient stability. The scaling factor is dynamically set as:

\[
\beta =
\begin{cases}
1, & \text{if update is stable} \\
0.1, & \text{if instability occurs} \\
0.01, & \text{for further correction}
\end{cases}
\]

This factor moderates updates to the dual coefficients and bias:

\[
\alpha_i \leftarrow \alpha_i + \beta \cdot \Delta \alpha_i, \quad b \leftarrow b + \beta \cdot \Delta b
\]

Here, \( \Delta \alpha_i \) and \( \Delta b \) denote the calculated updates for the dual coefficients and the bias term, respectively. This mechanism is used only during the training phase and helps stabilize convergence when large margin shifts or noisy samples are detected.

\begin{algorithm}[H]
\caption{Hybrid Kernel Evaluation with Conservative Scaling}
\label{alg:hybrid_kernel}
\KwIn{Normalized test sample \( X' \), support vectors \( \{SV_j\}_{j=1}^{k} \), polynomial kernel \( K_p \), RBF kernel \( K_r \), kernel weights \( \lambda_1, \lambda_2 \), dual coefficients \( \alpha_j \), scaling factor \( \beta \)}
\KwOut{Encrypted decision score \( D \)}

Initialize decision score: \( D \leftarrow 0 \)\;

\For{each support vector \( SV_j \)}{
  Evaluate polynomial kernel:\newline
  \hspace*{2em} \( K_p^j \leftarrow K_p(X', SV_j) \)\;

  Evaluate RBF kernel:\newline
  \hspace*{2em} \( K_r^j \leftarrow K_r(X', SV_j) \)\;

  Compute hybrid kernel output:\newline
  \hspace*{2em} \( K^j \leftarrow \lambda_1 \cdot K_p^j + \lambda_2 \cdot K_r^j \)\;

  Apply conservative scaling:\newline
  \hspace*{2em} \( \tilde{\alpha}_j \leftarrow \alpha_j + \beta \cdot \Delta \alpha_j \)\;

  Update decision score:\newline
  \hspace*{2em} \( D \leftarrow D + K^j \cdot \tilde{\alpha}_j \)\;
}
\Return \( D \)\;
\end{algorithm}

\subsection{Encrypted Inference}
After training, the PP-FinTech model performs encrypted inference, classifying test samples without ever decrypting them. We use the CKKS scheme with a 128-bit security level \text{(HEStd\_128\_classic)}, a ring dimension of 32,768, a modulus degree of 16,384, and the scaling factor \(\Delta\) of \(2^{20}\). A three-level modulus chain supports the multiplicative depth needed for kernel evaluation while balancing precision, performance, and noise growth. To manage noise from ciphertext multiplications, rescaling is applied after each operation. Parameter settings were selected to ensure computations remain within the noise budget, preventing decryption errors. The inference phase consists of four steps: data encryption, homomorphic kernel evaluation, adaptive thresholding, and result decryption.

\subsubsection{Data Encryption and HE Kernel Evaluation}
In the data encryption phase, data samples and model parameters are encrypted. For a given data sample \(X'\), each feature is individually encrypted using 
\[
C(X'_i) = \text{Enc}(X'_i)
\]

where \(C(X'_i)\) is the encrypted representation of the feature.The encrypted test sample is then represented as a vector of ciphertexts:

\[
C(X') = \{ C(X'_1), C(X'_2), ..., C(X'_n) \}
\]
Encrypted samples are processed using homomorphic operations to evaluate the SVM decision function without decryption. The encrypted decision score is computed as:

\[
C(S') = \sum_{j} C(\alpha_j) \cdot C(K(X', SV_j)) + C(b)
\]
The hybrid kernel output for each support vector is computed as:
\[
C(K(X', SV_j)) = \text{Enc}(K(X', SV_j))
\]

\subsubsection{SIMD-Based Encrypted Matrix Multiplication}
To boost efficiency, we leverage SIMD techniques for parallel computation within a single homomorphic operation~\cite{mustafa2024mimd}. Each test sample is packed into one ciphertext, enabling homomorphic dot products:

\[
C(D_j) = \sum_{i=1}^{n} C(X'_i) \cdot C(SV_{j,i})
\]

Where \(C(D_j)\) represents the encrypted dot product between the test sample (\(X'_i\)) and the corresponding support vector. For multiple test samples, this extends naturally to matrix form for batch processing.:
\noindent\[
\setlength{\abovedisplayskip}{10pt}
\setlength{\belowdisplayskip}{10pt}
C(D) = C(X') \cdot C(SV)^T
\]
which can be expanded as:
\[
\resizebox{\linewidth}{!}{$
C(D) = \left[
\begin{array}{ccc}
\sum\limits_{i=1}^{n} C(X'_{1,i}) \cdot C(SV_{1,i}) & \cdots & \sum\limits_{i=1}^{n} C(X'_{1,i}) \cdot C(SV_{k,i}) \\
\sum\limits_{i=1}^{n} C(X'_{2,i}) \cdot C(SV_{1,i}) & \cdots & \sum\limits_{i=1}^{n} C(X'_{2,i}) \cdot C(SV_{k,i}) \\
\vdots & \ddots & \vdots \\
\sum\limits_{i=1}^{n} C(X'_{m,i}) \cdot C(SV_{1,i}) & \cdots & \sum\limits_{i=1}^{n} C(X'_{m,i}) \cdot C(SV_{k,i})
\end{array}
\right]
$}
\]

To improve performance, SIMD batching was applied during encrypted matrix multiplications, enabling parallel inference over multiple samples within a single ciphertext. This optimization reduced per-sample latency and memory usage. While we did not benchmark SIMD separately, qualitative observations confirmed reduced total inference time, particularly when processing batches of test samples.

\subsubsection{Adaptive Threshold for Secure Classification}
Our model leverages adaptive thresholding which dynamically adjusts the decision boundary to improve classification results, and is computed as:
\[
\theta = \lambda_1 \cdot \mu + \frac{\lambda_2}{\sigma}
\]
where \(\mu\) represents the mean value of encrypted decision scores across multiple test samples, and \(\sigma\) outlines the variations in encrypted classification outputs. The weights \( \lambda_1 = 0.5 \) and \( \lambda_2 = 0.1 \) were determined empirically and fixed  to maintain consistent classification behavior across encrypted batches. 
This approach is, to our knowledge, among the first to embed adaptive thresholding within a CKKS-encrypted SVM pipeline, offering improved classification stability under encrypted inference conditions.

\begin{algorithm}[H]
\caption{Encrypted Inference with Adaptive Thresholding}
\label{alg:encrypted_inference}
\KwIn{Encrypted test sample \( \text{Enc}(X') \), encrypted support vectors \( \text{Enc}(SV_j) \), dual coefficients \( \alpha_j \), hybrid kernel weights \( \lambda_1, \lambda_2 \), model bias \( b \)}
\KwOut{Predicted class label (in encrypted form or after decryption)}

Initialize encrypted decision score: \( \text{Enc}(D) \leftarrow 0 \)\;

\For{each support vector \( SV_j \)}{
  Compute encrypted polynomial kernel:\newline
  \hspace*{2em} \( \text{Enc}(K_p^j) \leftarrow K_p(\text{Enc}(X'), \text{Enc}(SV_j)) \)\;
  
  Compute encrypted RBF kernel:\newline
  \hspace*{2em} \( \text{Enc}(K_r^j) \leftarrow K_r(\text{Enc}(X'), \text{Enc}(SV_j)) \)\;
  
  Compute encrypted hybrid kernel:\newline
  \hspace*{2em} \( \text{Enc}(K^j) \leftarrow \lambda_1 \cdot \text{Enc}(K_p^j) + \lambda_2 \cdot \text{Enc}(K_r^j) \)\;
  
  Multiply with plaintext coefficient:\newline
  \hspace*{2em} \( \text{Enc}(S_j) \leftarrow \alpha_j \cdot \text{Enc}(K^j) \)\;
  
  Accumulate score:\newline
  \hspace*{2em} \( \text{Enc}(D) \leftarrow \text{Enc}(D) + \text{Enc}(S_j) \)\;
}

Add encrypted bias:\newline
\hspace*{2em} \( \text{Enc}(D) \leftarrow \text{Enc}(D) + \text{Enc}(b) \)\;

Decrypt decision score:\newline
\hspace*{2em} \( D \leftarrow \text{Dec}(\text{Enc}(D)) \)\;

Compute adaptive threshold:\newline
\hspace*{2em} \( \theta = \lambda_1 \cdot \mu + \lambda_2 / \sigma \)\;

\Return \( \text{sign}(D - \theta) \)\;
\end{algorithm}

\subsubsection{Secure Decryption and Classification}
After encrypted inference, the classification results are sent to the client for decryption using their private key:
\[
S' = \text{Dec}(C(S'))
\]
The classification label is then determined with adaptive thresholds:
\[
y' = 
\begin{cases} 
1, & \text{if } S' > \theta \\
0, & \text{otherwise}
\end{cases}
\]
\section{Results and Discussion}
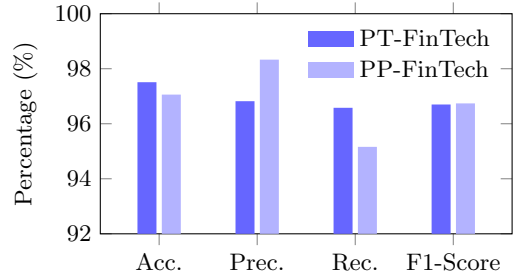
\begin{figure}[h]
\centering
\begin{tikzpicture}
\begin{axis}[
    ybar,
    bar width=0.25cm,
    width=7cm,
    height=4.5cm,
    ylabel={Percentage (\%)},
    symbolic x coords={Acc., Prec., Rec., F1-Score},
    xtick=data,
    ymin=92,
    ymax=100,
    enlarge x limits=0.2,
    legend style={
        at={(0.98,0.98)},
        anchor=north east,
        draw=none,
        fill=none,
        font=\small
    },
    legend cell align={left}
]
\addplot+[ybar, area legend, fill=blue!60, draw opacity=0] coordinates
{(Acc.,97.51) (Prec.,96.82) (Rec.,96.58) (F1-Score,96.70)};
\addplot+[ybar, area legend, fill=blue!30, draw opacity=0] coordinates
{(Acc.,97.06) (Prec.,98.33) (Rec.,95.16) (F1-Score,96.74)};
\legend{PT-FinTech,PP-FinTech}
\end{axis}
\end{tikzpicture}
\caption{Performance Comparison Between PT-Fintech and PP-FinTech}
\label{fig:performance_comparison}
\end{figure}

In this section, we evaluate the performance of four models:

\begin{itemize}
    \item \textbf{PT-Linear}: A baseline plaintext model that uses a standard linear SVM kernel. This model serves as a reference point for performance without encryption or advanced kernel techniques..
    \item \textbf{PP-Linear}: An encrypted version of the linear SVM model, implemented using the CKKS homomorphic encryption scheme which performs inference directly on encrypted data.
    \item \textbf{PT-FinTech}: A plaintext hybrid-kernel model that combines Polynomial and Radial Basis Function (RBF) kernels to capture more complex, nonlinear relationships in the data.
    \item \textbf{PP-FinTech}: An encrypted hybrid-kernel SVM model combining Polynomial and RBF kernels using CKKS.
\end{itemize}
 Fig.~\ref{fig:performance_comparison} compares only two models — PT-Fintech and PP-Fintech — to highlight the effect of encryption on the performance of the hybrid-kernel SVM.
We assessed the models using four key metrics: accuracy, precision, recall, and F1-score, defined by Equations (1)–(4):

\begin{align}
\text{Accuracy} &= \frac{TP + TN}{TP + TN + FP + FN} \\
\text{Precision} &= \frac{TP}{TP + FP} \\
\text{Recall} &= \frac{TP}{TP + FN} \\
\text{F1-Score} &= 2 \cdot \frac{\text{Recall} \cdot \text{Precision}}{\text{Recall} + \text{Precision}}
\end{align}

Here, TP and TN denote correctly predicted positive and negative instances, while FP and FN represent misclassified negatives and positives, respectively ~\cite{reddy2022classification}. Fig. ~\ref{fig:performance_comparison} illustrates a comparative analysis of accuracy, precision, recall, and F1-score between our proposed PP-FinTech and the baseline model (PT-FinTech). This evaluation aims to assess the impact of integrating CKKS-based FHE into our model. The results show that PP-FinTech achieved a precision of 98.33\%, slightly higher than PT-FinTech's 96.82\%. Conversely, PT-FinTech attained a marginally higher accuracy of 97.51\%, compared to 97.06\% for PP-FinTech. Recall values were 96.58\% for PT-FinTech and 95.16\% for PP-FinTech. The F1-score was nearly identical across both models, with 96.74\% for PP-FinTech and 96.70\% for PT-FinTech. While minor differences are observed, they are not statistically significant and fall within the expected margin of variation. These findings confirm that the encrypted PP-FinTech model delivers comparable predictive performance to its non-encrypted counterpart, despite the computational noise introduced by FHE. The slightly higher precision of PP-FinTech may reflect improved robustness in positive classifications, though such differences should be interpreted cautiously. Overall, the results justify the use of FHE in privacy-preserving inference without compromising model effectiveness. A detailed comparison across all evaluation metrics is presented in Table~\ref{tab:baseline_comparison}.

\subsection{ROC Curve}
\begin{figure}[h]
\centering
\begin{tikzpicture}
\begin{axis}[
    width=0.65\linewidth,
    height=5cm,
    grid=major,
    xlabel=False Positive Rate (FPR),
    ylabel=True Positive Rate (TPR),
    xmin=0, xmax=1,
    ymin=0, ymax=1,
    axis equal image,
    legend pos=south east,
    legend style={font=\footnotesize},
]

\addplot[dashed, gray] coordinates {(0,0)(1,1)};
\addlegendentry{Random Guess}

\addplot[
    color=blue,
    thick,
    mark=*,
    mark options={scale=0.6}
]
coordinates {
    (0,0)
    (0.02, 0.70)
    (0.04, 0.82)
    (0.06, 0.89)
    (0.08, 0.93)
    (0.10, 0.9506)
    (1,1)
};
\addlegendentry{PT-FinTech}

\addplot[
    color=red,
    thick,
    mark=square*,
    mark options={scale=0.6}
]
coordinates {
    (0,0)
    (0.015, 0.72)
    (0.03, 0.84)
    (0.05, 0.90)
    (0.07, 0.94)
    (0.09, 0.9516)
    (1,1)
};
\addlegendentry{PP-FinTech}

\end{axis}
\end{tikzpicture}
\caption{ROC Curve Comparison}
\label{fig:roc_curve}
\end{figure}
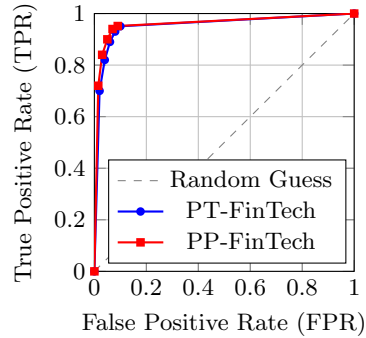

In addition to standard evaluation metrics, we examined the Receiver Operating Characteristic (ROC) curve to further evaluate model classification performance. The ROC curve plots the True Positive Rate (TPR) against the False Positive Rate (FPR) across varying classification thresholds, providing insight into the model's ability to distinguish between classes  ~\cite{carrington2022deep} ~\cite{miao2022precision}. A key summary metric is the Area Under the Curve (AUC), where higher values indicate better discriminatory power ~\cite{lee2021induction}. Fig. ~\ref{fig:roc_curve} shows both models exhibit strong performance indicating high TPR and low FPR across thresholds. Notably, the encrypted PP-FinTech model maintains a performance nearly identical to its plaintext counterpart, suggesting that encryption introduces minimal degradation in classification effectiveness.

\subsection{Computational Overhead}
We also evaluated computation time across all four models to measure efficiency. As shown in Table~\ref{tab:baseline_comparison}, encryption adds noticeable latency, with PP-FinTech showing higher inference time than PT-FinTech due to the overhead of FHE. However, this is mitigated by SIMD batching and optimized CKKS parameters. Despite the added cost, PP-FinTech maintains practical latency, making it viable for privacy-sensitive financial applications with moderate throughput needs.
\enlargethispage{0.09cm}
\begin{table*}[h]
\centering
\small
\caption{Baseline Comparison of Linear and Hybrid SVM Models}
\label{tab:baseline_comparison}
\begin{tabular}{|l|c|c|c|c|c|c|}
\hline
\textbf{Model} & \textbf{Encrypted} & \textbf{Acc. (\%)} & \textbf{Pre. (\%)} & \textbf{Rec. (\%)} & \textbf{F1-Score (\%)}  & \textbf{Time (ms)} \\
\hline
PT-Linear   & No  & 90.53  & 88.26 &  89.03 &  88.58 & 0.34      \\
PP-Linear   & Yes & 89.04  & 87.05 & 88.01 & 87.48 & 1.75      \\
PT-FinTech  & No  & 97.51  & 96.82 & 96.58 & 96.70 & 2.91       \\
PP-FinTech  & Yes & 97.06  & 98.33 & 95.16 & 96.74 & 44.9      \\
\hline
\end{tabular}
\end{table*}

\subsubsection{Runtime Breakdown}
\begin{figure}[ht]
\centering

\begin{subfigure}[t]{0.48\linewidth}
\centering
\begin{tikzpicture}
\begin{axis}[
    ybar,
    bar width=7pt,
    width=0.9\linewidth,
    height=5cm,
    ymin=0, ymax=25,
    ylabel={\textbf{Time (ms)}},
    ylabel style={font=\bfseries\footnotesize},
    symbolic x coords={Enc, Kernel, Thresh, Dec},
    nodes near coords align={vertical},
    every node near coord/.append style={black},
    enlarge x limits=0.2,
    grid=major,
    tick label style={font=\small},
    x tick label style={yshift=120pt, rotate=65, anchor=east, font=\small}
]
\addplot+[fill=blue!55] coordinates {
    (Enc,10)
    (Kernel,23.5)
    (Thresh,7)
    (Dec,4.4)
};
\end{axis}

\end{tikzpicture}
\caption{Runtime Breakdown}
\label{fig:runtime}
\end{subfigure}
\hfill
\begin{subfigure}[t]{0.48\linewidth}
\centering
\begin{tikzpicture}
\begin{axis}[
    ybar,
    bar width=7pt,
    width=0.9\linewidth,
    height=5cm,
    ymin=0, ymax=130,
    ylabel={\textbf{Noise (bits)}},
    symbolic x coords={Enc, Kernel, Thresh, Dec},
    xtick=data,
    nodes near coords align={vertical},
    every node near coord/.append style={black},
    enlarge x limits=0.2,
    grid=major,
    tick label style={font=\small},
    xticklabel style={yshift=120pt, rotate=65, anchor=east, font=\small}
]
\addplot+[fill=blue!55] coordinates {
    (Enc,120)
    (Kernel,90)
    (Thresh,60)
    (Dec,45)
};
\end{axis}
\end{tikzpicture}
\caption{Noise Budget Analysis}
\label{fig:nsa}
\end{subfigure}

\caption{Empirical analysis of runtime and noise behavior during encrypted inference.}
\label{fig:runtime_noise_combined}
\end{figure}
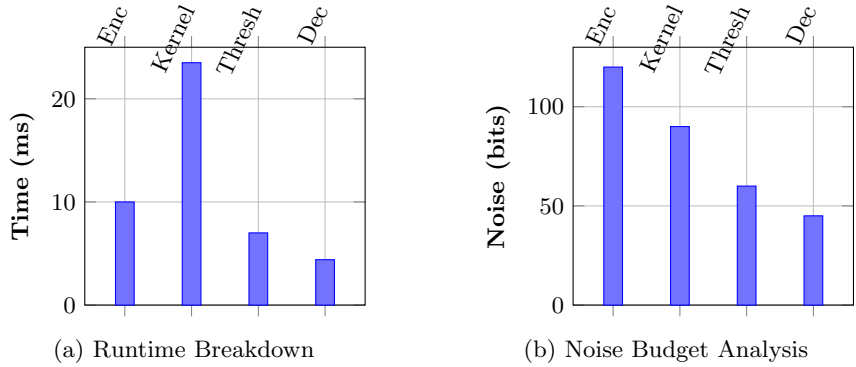

To analyze the performance of the encrypted inference pipeline, we profiled each stage to identify its contribution to the total latency. As shown in Fig. ~\ref{fig:runtime}, hybrid kernel evaluation was the most time-consuming step, averaging 23.5~ms per sample. Encryption and adaptive thresholding took 10~ms and 7~ms respectively, while decryption was relatively fast at 4.4~ms. This results in a total average inference time of 44.9~ms per test sample, demonstrating the efficiency of our pipeline under FHE constraints.

\subsection{Noise Budget Analysis}
During encrypted inference, each homomorphic operation in CKKS contributes to cumulative noise growth. To visualize this effect, Fig. ~\ref{fig:nsa} presents the remaining noise budget across the main stages of our inference pipeline: initial encryption, hybrid kernel evaluation, adaptive thresholding, and the final output before decryption. The noise budget begins at approximately 120 bits and decreases with each multiplication and rescaling step, reaching around 45 bits before decryption. This confirms that our chosen CKKS parameters provide sufficient multiplicative depth while keeping the noise within safe limits, ensuring accurate decryption at the end of inference.

\subsection{Scalability Evaluation}

\begin{figure}[ht]
\centering
\begin{tikzpicture}
\begin{axis}[
    width=0.75\linewidth,
    height=6cm,
    xlabel={\textbf{Batch Size (samples)}},
    ylabel={\textbf{Total Inference Time (ms)}},
    ymin=0, ymax=5,
    ytick={0,1,2,3,4,5},
    xtick=data,
    grid=major,
    enlargelimits=0.1,
    tick label style={font=\small},
    label style={font=\bfseries\small},
]
\addplot+[mark=*, color=blue, thick] coordinates {
    (10, 0.449)
    (25, 1.123)
    (50, 2.245)
    (75, 3.368)
    (100, 4.490)
};
\end{axis}
\end{tikzpicture}
\caption{Simulated scalability: Total inference time across different batch sizes for the PP-FinTech model.}
\label{fig:scaling_graph}
\end{figure}
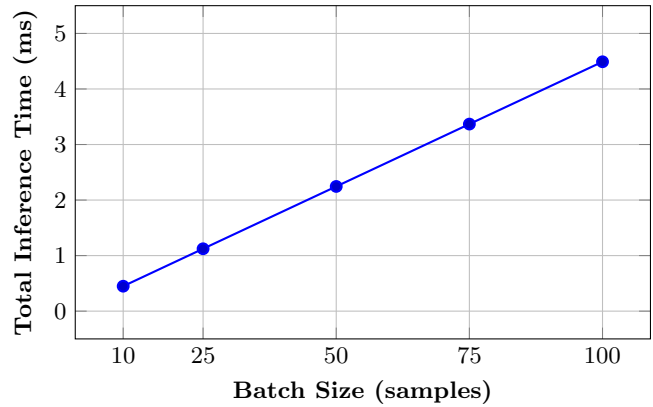

To assess the scalability of our encrypted inference system, we simulated batch inference using the observed per-sample latency of approximately 44.9~ms. As shown in Fig. ~\ref{fig:scaling_graph}, the total inference time increases linearly with the number of test samples, confirming predictable scalability under larger workloads. This stable per-sample latency is attributed to the efficiency of SIMD batching in our encrypted pipeline, which enables parallel processing of multiple encrypted operations.

\section{Conclusion}

This paper presented a privacy-preserving classification model that integrates SVM with the CKKS homomorphic encryption scheme for secure credit scoring. By employing a hybrid kernel and adaptive thresholding, the model effectively handles non-linear patterns while enhancing robustness. Despite the added encryption overhead, it delivers performance comparable to its plaintext counterpart with realistic latency, making it practical for privacy-sensitive financial applications. Future work will explore extensions to deep learning and real-time datasets to further improve scalability and efficiency.

\bibliographystyle{IEEEtran}
\bibliography{references}

\end{document}